\documentclass{article}
\usepackage[english]{babel} 

\usepackage{graphicx}

\usepackage{color}

\begin{document}

\newcommand{\rum}{\rule{0.5pt}{0pt}}
\newcommand{\rub}{\rule{1pt}{0pt}}
\newcommand{\rim}{\rule{0.3pt}{0pt}}
\newcommand{\numtimes}{\mbox{\raisebox{1.5pt}{${\scriptscriptstyle \rum\times}$}}}
\newcommand{\numtimess}{\mbox{\raisebox{1.0pt}{${\scriptscriptstyle \rum\times}$}}}
\newcommand{\Boldsq}{\vbox{\hrule height 0.7pt
\hbox{\vrule width 0.7pt \phantom{\footnotesize T}%
\vrule width 0.7pt}\hrule height 0.7pt}}
\newcommand{\two}{$\raise.5ex\hbox{$\scriptstyle 1$}\kern-.1em/
\kern-.15em\lower.25ex\hbox{$\scriptstyle 2$}$}

\renewcommand{\refname}{References}
\renewcommand{\tablename}{\small Table}
\renewcommand{\figurename}{\small Fig.}
\renewcommand{\contentsname}{Contents}

\begin{center}
{\Large\bf 
Dynamical 3-Space Gravity Theory: Effects on Polytropic Solar Models\rule{0pt}{13pt}}\par

\bigskip

Richard D. May \\
{\small\it  School of Chemical and Physical  Sciences, Flinders University,
Adelaide 5001, Australia\rule{0pt}{15pt}}\\
\raisebox{+1pt}{\footnotesize E-mail: Richard.May@flinders.edu.au}\par

\bigskip

Reginald T. Cahill \\ 
{\small\it  School of Chemical and Physical  Sciences, Flinders University,
Adelaide 5001, Australia\rule{0pt}{15pt}}\\
\raisebox{+1pt}{\footnotesize E-mail: Reg.Cahill@flinders.edu.au}\par

\bigskip

{\small\parbox{11cm}{%
Numerous experiments and observations have confirmed the existence of a dynamical 3-space,  detectable directly by light-speed anisotropy experiments, and indirectly by means of novel gravitational effects, such as bore hole $g$ anomalies, predictable black hole masses, flat spiral-galaxy rotation curves, and the expansion of the universe, all without  dark matter and dark energy.  The dynamics for this 3-space follows from a unique generalisation of Newtonian gravity, once that is cast into a velocity formalism. This new theory of gravity is applied to the  sun to compute new density, pressure and temperature profiles, using  polytrope modelling of the equation of state for the matter.  These results  should be applied to a re-analysis of solar neutrino production, and to stellar evolution in general.

 \rule[0pt]{0pt}{0pt}}}\medskip
\end{center}

\setcounter{section}{0}
\setcounter{equation}{0}
\setcounter{figure}{0}
\setcounter{table}{0}

\markboth{May and Cahill.  Dynamical 3-Space Gravity Theory: Effects on Polytropic Solar Models }{\thepage}
\markright{May and Cahill.  Dynamical 3-Space Gravity Theory: Effects on Polytropic Solar Models  }

\setcounter{section}{0}
\setcounter{equation}{0}
\setcounter{figure}{0}
\setcounter{table}{0}

\tableofcontents

\section{Introduction}
It has been discovered that Newton's theory of gravity \cite{Newton} missed a significant dynamical process, and a uniquely determined  generalisation to include this process has resulted in the explanation of numerous gravitational anomalies, such as
 bore hole $g$ anomalies, predictable black hole masses, flat spiral-galaxy rotation curves, and the expansion of the universe, all without  dark matter and dark energy \cite{Book,Review,Paradigm}. This theory of gravity arises from  the dynamical 3-space, described by a dynamical velocity field,  when the Schr\"{o}dinger equation is generalised to  take account of the propagation of quantum matter in the dynamical 3-space. So  gravity is now an emergent phenomenon, together with the equivalence principle. 
 
 The dynamical 3-space has been directly observed using various light-speed anisotropy experiments, dating from the 1st detection by Michelson and Morley in 1887 \cite{MMCK,MMC}, giving a speed in excess of 300km/s, after re-calibrating the gas-mode interferometer for actual length contraction effects, to the latest using spacecraft earth-flyby Doppler shift data \cite{CahillNASA}. Overall these experiments reveal that relativistic effects are caused by the absolute motion of rods and clocks wrt the dynamical 3-space, essentially Lorentzian Relativity (LR), rather than the Special Relativity(SR)  formalism, which has recently been shown by means of an exact change of space and time variables, to be equivalent to Galilean Relativity \cite{CahillMink}.
 
 Here we apply the new gravity theory to the internal dynamics of the sun, and compute  new density, pressure and temperature profiles, using  the polytrope model for the equation of state of the matter.  These results  should then be applied to a re-analysis of neutrino production \cite{Bahcall}.
 In general the Newtonian-gravity based standard model of stellar evolution  also needs re-examination. 

\section{Dynamical 3-Space}

Newton's inverse square law of gravity  has the differential form
\begin{equation}
\nabla.{\bf g}=-4\pi G\rho,  \mbox{\ \ \  } \nabla \times {\bf g} = {\bf 0}, 
\label{eqn:Newton}\end{equation}
for the  acceleration field ${\bf g}({\bf r},t)$, assumed to be fundamental and existing within Newton's model of space,  which is Euclidean, static, and unobservable.
Application of this to spiral galaxies and the expanding universe has lead to many problems, including, in part, the need to invent dark energy and dark matter\footnote{The Friedmann equations for the expanding universe follow trivially from (\ref{eqn:Newton}), as shown in \cite{Paradigm}, but then needs ``dark matter" and ``dark energy" to fit the cosmological data.}.  However (\ref{eqn:Newton}) has a unique generalisation that resolves these and other problems.   In terms of a velocity field ${\bf v}({\bf r},t)$ (\ref{eqn:Newton}) has an equivalent form \cite{Book,Review}
\begin{equation}
\nabla.\left(\frac{\partial {\bf v} }{\partial t}+ ({\bf v}.{\bf \nabla}){\bf v}\right)
=-4\pi G\rho,  \mbox{\ \ \  } \nabla \times {\bf v} = {\bf 0}, 
\label{eqn:Newtonv} \end{equation}
where now
\begin{equation}
{\bf g}=\frac{\partial{\bf v}}{\partial t}+({\bf v}.\nabla){\bf v},
\label{eqn:acceln}\end{equation}
is the well-known Galilean covariant Euler acceleration of the substratum that has velocity  ${\bf v}({\bf r},t)$.
Because of the covariance of ${\bf g}$ under a change of the spatial coordinates only relative internal velocities have an ontological existence - the coordinates ${\bf r}$  then merely define a mathematical embedding space.

 We give a brief review of the concept and mathematical formalism of a dynamical flowing 3-space, as this is often confused with the older dualistic space and aether ideas, wherein some particulate aether is located and  moving through an unchanging Euclidean space - here both the space and the aether were viewed as being ontologically real.  The dynamical 3-space is different: here we have only a dynamical 3-space, which at a small scale is a quantum foam system without dimensions and described by fractal  or nested homotopic mappings \cite{Book}.  This quantum foam is not embedded in any space -  the quantum foam is all there is, and any metric properties are intrinsic properties solely of that quantum foam. At a macroscopic level the quantum foam is described by a velocity field ${\bf v}({\bf r},t)$, where ${\bf r}$ is merely a $[3]$-coordinate within an embedding space. This embedding space has no ontological existence - it is merely used to (i) record that the quantum foam has, macroscopically, an effective dimension of 3, and (ii) to relate other phenomena also described by fields, at the same point  in the quantum foam.  The dynamics for this 3-space is easily determined by the requirement that observables be independent of the embedding choice, giving, for zero-vorticity  dynamics and for a flat embedding space,  and preserving the inverse square law outside of spherical masses, at least in the usual cases, such as planets,
 \begin{equation}
\nabla.\left(\frac{\partial {\bf v} }{\partial t}+ ({\bf v}.{\bf \nabla}){\bf v}\right)+
\frac{\alpha}{8}\left((tr D)^2 -tr(D^2)\right)=-4\pi G\rho, \nonumber \end{equation}
\begin{equation}
 \nabla\times {\bf v}={\bf 0},  \mbox{\  \  \   }
 D_{ij}=\frac{1}{2}\left(\frac{\partial v_i}{\partial x_j}+
\frac{\partial v_j}{\partial x_i}\right),
\label{eqn:3spacedynamics}\end{equation} 
where $\rho({\bf r},t)$ is the matter and EM energy densities, expressed as an effective matter density.  Borehole $g$ measurements and astrophysical black hole data has shown that $\alpha\approx1/137$ is  the fine structure constant to within observational errors  \cite{Book, Review,Schrod}.  For a quantum system with mass
$m$ the Schr\"{o}dinger equation is  uniquely generalised   \cite{Schrod} with the new terms required to maintain that the motion is intrinsically wrt the 3-space, and not wrt the embedding space,  and that the time evolution is unitary:
\begin{equation}
i\hbar\frac{\partial \psi({\bf r},t)}{\partial t}  =-\frac{\hbar^2}{2m}\nabla^2\psi({\bf r},t)-i\hbar\left({\bf
v}.\nabla+\frac{1}{2}\nabla.{\bf v}\right) \psi({\bf r},t).
\label{eqn:Schrod}\end{equation}
The space and time coordinates $\{t,x,y,z\}$ in (\ref{eqn:3spacedynamics}) and (\ref{eqn:Schrod}) ensure that  the separation of a deeper and unified process into different classes of phenomena - here a dynamical 3-space (quantum foam) and a quantum matter system, is properly tracked and connected. As well the same coordinates may be used by an observer to also track the different phenomena.  However it is important to realise that these coordinates have no ontological significance - they are not real.  The velocities ${\bf v}$ have no ontological or absolute meaning relative to this coordinate system - that is in fact how one arrives at the form in  (\ref{eqn:Schrod}), and so the ``flow" is always relative to the internal dynamics of the 3-space. 
A quantum wave packet propagation analysis of (\ref{eqn:Schrod}) gives  the acceleration induced by wave refraction to be \cite{Schrod}
\begin{eqnarray}
{\bf g}=\frac{\partial{\bf v}}{\partial t}+({\bf v}.\nabla){\bf v}+
(\nabla\times{\bf v})\times{\bf v}_R,\nonumber \\ {\bf v}_R({\bf r}_o(t),t) ={\bf v}_o(t) - {\bf v}({\bf r}_o(t),t),
\label{eqn:accelnb}\end{eqnarray}
where ${\bf v}_R$ is the velocity of the wave packet relative to the local 3-space, and where ${\bf v}_o$ and ${\bf r}_o$ are the velocity and position relative to the observer, and the last term in (\ref{eqn:accelnb}) generates the Lense-Thirring effect as a vorticity driven effect.  Together (\ref{eqn:3spacedynamics}) and (\ref{eqn:accelnb})  amount to the derivation of gravity as a quantum effect,  explaining  both the  equivalence principle ($\bf g$ in (\ref{eqn:accelnb}) is independent of $m$) and the Lense-Thirring effect. Overall we see, on ignoring vorticity effects, that
\begin{equation}
\nabla.{\bf g}=-4\pi G\rho-4\pi G\rho_{DM},
\label{eqn:NGplus}\end{equation}
where
\begin{equation}
\rho_{DM} =\frac{\alpha}{32\pi G}\left((tr D)^2 -tr(D^2)\right).
\label{eqn:darkmatter}\end{equation}
This is Newtonian gravity but with the extra dynamical term which has been used to define an effective ``dark matter" density. This is not  real matter, of any form, but is the matter density needed within Newtonian gravity to explain the flat rotation curves of spiral galaxies, large light bending and lensing effects from galaxies, and other effects.  Here, however, it is purely a space self-interaction effect.  This new dynamical effect also explains  the bore hole $g$ anomalies,  and the black hole ``mass spectrum". Eqn.(\ref{eqn:3spacedynamics}), even when $\rho=0$, has an expanding universe Hubble solution that fits the recent supernovae data in a parameter-free manner without requiring ``dark matter" nor ``dark energy", and without the accelerating expansion artifact \cite{Paradigm}. However (\ref{eqn:NGplus}) cannot be entirely expressed in terms  of ${\bf g}$ because the fundamental dynamical variable is $\bf v$. The role of  (\ref{eqn:NGplus}) is to reveal that if we analyse gravitational phenomena we will usually find that the matter density $\rho$ is insufficient to account for the observed ${\bf g}$. Until recently this failure of Newtonian gravity has been explained away as being caused by some unknown and undetected ``dark matter" density.  Eqn.(\ref{eqn:NGplus}) shows that to the contrary it is a dynamical property of 3-space itself. 
Significantly the quantum matter 3-space-induced  `gravitational' acceleration in (\ref{eqn:accelnb}) also follows from maximising the elapsed proper time wrt the wave-packet trajectory ${\bf r}_o(t)$, see \cite{Book},
\begin{equation}
\tau=\int dt \sqrt{1-\frac{{\bf v}^2_R({\bf r}_o(t),t)}{c^2}},
\label{eqn:propertime}\end{equation}
and then taking the limit $v_R/c \rightarrow 0$. This shows that (i) the matter `gravitational' geodesic is a quantum wave refraction effect, with the trajectory determined by a Fermat maximised proper-time  principle, and (ii) that quantum systems undergo a local time dilation effect.  Significantly the time dilation effect in (\ref{eqn:propertime}) involves matter motion wrt the dynamical 3-space, and not wrt the observer, and so distinguishing LR from SR. A full derivation of 
(\ref{eqn:propertime}) requires the generalised Dirac equation, with the replacement $\partial/\partial t \rightarrow  \partial/\partial t +{\bf v}\cdot\nabla$, as in (\ref{eqn:Schrod}).  In differential form (\ref{eqn:propertime})
becomes
\begin{equation}
d\tau^2=g_{\mu\nu}dx^\mu dx^\nu=dt^2-\frac{1}{c^2}(d{\bf r}(t)-{\bf v}({\bf r}(t),t)dt)^2,
\label{eqn:PGmetric}\end{equation}
which introduces a curved spacetime metric $g_{\mu\nu}$ that emerges from (\ref{eqn:3spacedynamics}). However this spacetime has no ontological significance - it is merely a mathematical artifact, and as such hides the underlying dynamical 3-space. This induced metric is not determined by the Einstein-Hilbert equations, which originated as a generalisation of Newtonian gravity, but without the knowledge that a dynamical 3-space had indeed been detected by Michelson and Morley in 1887 by detecting light speed anisotropy.  In special circumstances, and with $\alpha=0$,  they do yield the same effective spacetime metric.  However the dynamics in (\ref{eqn:3spacedynamics}) is more general, as noted above, and has passed more tests.

\section{New Gravity Equation for a Spherically Symmetric System}

For the  case of zero vorticity the matter acceleration in (\ref{eqn:accelnb}) gives
\begin{equation}
\vec g(\vec r, t)=\frac{\partial \vec v}{\partial t} + \frac{\vec \nabla \vec v^2}{2}
\end{equation}
For a time independent flow  we introduce a generalised  gravitational potential, which gives a microscopic explanation for that potential,
\begin{equation}
\Phi(\vec r)=-\frac{\vec v^2}{2}.
\label{eqn:PotentialDef}
\end{equation}
For the case of a spherically symmetric and time independent inflow we set \(\vec{v}(\vec{r}, t)=-\vec{\hat{r}} v(r)\) then (\ref{eqn:3spacedynamics}) becomes, with $v^\prime=dv/dr$,
\begin{equation}
\frac{\alpha}{2 r}\left(\frac{v^2}{2r}+v v' \right) + \frac{2}{r} v v' + (v')^2 + v v''=-4 \pi G \rho
\end{equation}
which can be written as
\begin{equation}
\frac{1}{r^2}\frac{d}{dr}\left(r^{2-\frac{\alpha}{2}}\frac{d}{dr}\left(r^{\frac{\alpha}{2}} \Phi\right)\right)=4 \pi G \rho
\label{eqn:Gravity}
\end{equation}
This form suggests that the new dynamics can be incorporated into the space metric, in that the 3-space $\alpha$-term appears to lead to a fractal dimension of $3-\alpha/2  = 2.996$, see \cite{Schrod}. The velocity flow description of space is completely equivalent to Newtonian gravity when the \( \alpha \) dependent term in (\ref{eqn:3spacedynamics}) is removed. In this case setting $\alpha=0$ reduces (\ref{eqn:Gravity}) to the Poisson equation of Newtonian gravity for the case of spherical symmetry.

\section{Solutions to New Gravity Equation for Non-Uniform Density}

The solutions to (\ref{eqn:Gravity}) for a uniform density distribution are published in \cite{Book}. For variable density $\rho(r)$ the exact solution to  (\ref{eqn:Gravity})  is\footnote{Eqn (\ref{eqn:Gravity})  also permits a $-\overline{\gamma}/r$ term in (\ref{eqn:RawGravitySol}).   However this is not valid, as the full  $[3]$ version of (\ref{eqn:Gravity}) would then involve a point mass at $r=0$, because $\nabla^2(1/|r|)=-4\pi \delta({\bf r})$, and  in (\ref{eqn:RawGravitySol}) all the mass is accounted for by $\rho(r)$. See \cite{Book}  for a detailed discussion.}
\begin{eqnarray}
\Phi(r)=
-  & \displaystyle {\frac{\beta}{r^\frac{\alpha}{2}} - \frac{G}{(1-\frac{\alpha}{2})r} \int_0^r 4 \pi s^2 \rho(s) ds}    \nonumber  \\
  -& \displaystyle {\frac{G}{(1-\frac{\alpha}{2})r^\frac{\alpha}{2}} \int_r^\infty 4 \pi s^{1+\frac{\alpha}{2}} \rho(s) ds} ,\label{eqn:RawGravitySol1}
\end{eqnarray}
When $\rho(r)=0$ for $r>R$, this becomes
\begin{equation}
\Phi(r)=\left\{
\begin{array}{l} 
     \displaystyle -  \frac{\beta}{r^\frac{\alpha}{2}} - \frac{G}{(1-\frac{\alpha}{2})r} \int_0^r 4 \pi s^2 \rho(s) ds \smallskip \\ 
     \displaystyle  - \frac{G}{(1-\frac{\alpha}{2})r^\frac{\alpha}{2}} \int_r^R 4 \pi s^{1+\frac{\alpha}{2}} \rho(s) ds , \mbox{\ \  } 0 < r \leq R \medskip\\
     \displaystyle - \frac{ \beta}{r^\frac{\alpha}{2}}- \frac{\gamma}{r} , \qquad r > R
\end{array}\right.
\label{eqn:RawGravitySol}
\end{equation}
where
\begin{equation}
\label{eqn:gamma}
\gamma= \frac{G}{(1-\frac{\alpha}{2})} \int_0^R 4 \pi s^2 \rho(s) ds =   \frac{GM}{(1-\frac{\alpha}{2})}
\end{equation}
 Here $M$ is the total matter mass, and $\beta$ is a free parameter. The term $\beta/r^{\alpha/2}$ describes an inflow singularity or ``black hole" with arbitrary strength. This is unrelated to the putative black holes of General Relativity. This corresponds to a primordial black hole.  As well the middle term in  (\ref{eqn:RawGravitySol})
also has a $1/r^{\alpha/2}$ inflow-singularity, but whose strength is mandated by the matter density, and is absent when $\rho(r)=0$ everywhere. This is a minimal ``black hole", and is present in all matter systems.  The  $\beta/r^{\alpha/2}$ term will produce a long range gravitational acceleration $g=\beta/r^{1+\alpha/2}$, as observed in spiral galaxies.  For the region outside the sun \((r>R)\) Keplerian orbits are known to well describe the motion of the planets within the solar system, apart from some small corrections, such as the Precession of the Perihelion of Mercury, which follow from relativistic effects from (\ref{eqn:propertime}).   Thus is  the case $\beta=0$, and the sun has only an  induced  `Minimal Attractor'.  These minimal black holes contribute to the external $g=K/r^2$ gravitational acceleration, through an effective mass
\begin{equation}
M_{BH}= \frac{M}{1-\frac{\alpha}{2}}-M=\frac{\alpha}{2} \frac{M}{1-\frac{\alpha}{2}}\approx \frac{\alpha}{2} M
\label{eqn:BHmass}\end{equation}
as previously reported \cite{Book}.  
These induced  black hole `` effective" masses have been detected in numerous  globular clusters and  spherical galaxies and  their predicted effective masses have been confirmed in some 19 such cases \cite{CahillBH2}. These gave the value  $\alpha =\approx 1/137$ \cite{CahillBH1}.   The induced black hole dynamics at the center of the sun is responsible for the new density, pressure and temperature profiles computed herein.

\section{Polytropic Models using Dynamical 3-Space Theory}
For a star to be in hydrostatic equilibrium  the inward force of gravity must match the net outward  effect of the pressure, 
\begin{equation}
\frac{dP}{dr}=-\frac{d\Phi}{dr} \rho
\label{eqn:HydroStatic}
\end{equation}
Here we use the polytrope modelling of the pressure-density equation of state. 
\begin{equation}
P=K \rho^{1+\frac{1}{n}}
\label{eqn:Polytrope}\end{equation}
where $n$ is the polytropic index, and $K$ is a constant. This was introduced by Lane and  Emden, and was extensively used by Chandrasekhar  \cite{ChandBook,Hansen,Horedt,KippBook}, but these analyses only  apply in  the case of Newtonian gravity.  The new theory of gravity  requires  a new treatment.

The polytropic relation between pressure and density (\ref{eqn:Polytrope})   gives
\begin{equation}
\frac{dP}{dr}=\frac{K (n+1)}{n} \rho^\frac{1}{n} \frac{d\rho}{dr}
\end{equation}
and (\ref{eqn:HydroStatic}) gives
\begin{equation}
\frac{d\Phi}{dr}=-\frac{K (n+1)}{n} \rho^{\frac{1}{n}-1} \frac{d\rho}{dr}
\end{equation} 
Integration  gives
\begin{equation}
\Phi=-K (n+1) \rho^{\frac{1}{n}} + C
\end{equation}

\begin{figure}
\hspace{30mm}\includegraphics[scale=0.6]{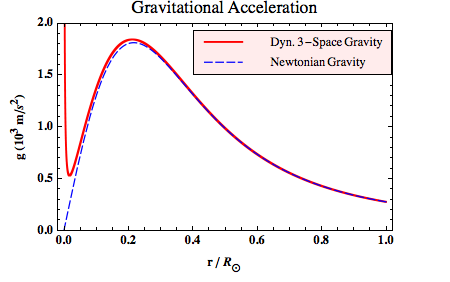} 

\hspace{30mm}\includegraphics[scale=0.6]{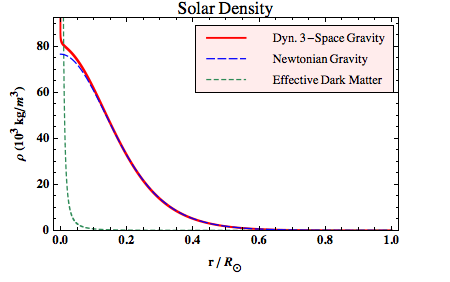} 
\caption{\small{Gravity and density plots for a polytropic model for the sun with \(n=3\). The effective dark matter distribution is shown in the density plot.}}
\label{fig:GravityDensity}
\end{figure}

 Here it will be useful to define the gravitational potential at the sun's surface  \(\Phi_R=\Phi(R)=C\) as the value of the integration constant, and so we obtain for the density
\begin{equation}
\rho=\left(\frac{\Phi_R - \Phi}{K (n+1)}\right)^n
\label{eqn:PolyCond}
\end{equation}

\begin{figure}
\hspace{30mm}\includegraphics[scale=0.6]{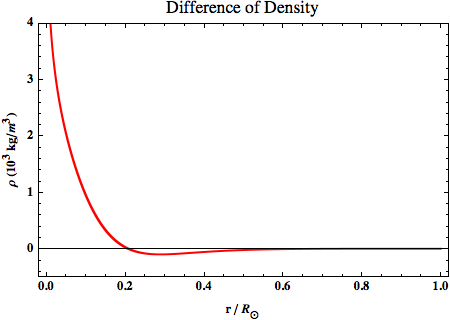} 

\hspace{30mm}\includegraphics[scale=0.6]{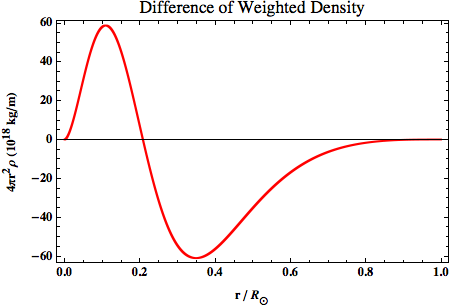} 
\caption{\small{Top graph shows difference in density $\rho(r)$ between new gravity and Newtonian modeling.  Bottom graph show the difference in weighted density  $4\pi r^2\rho(r)$. }}
\label{fig:DensityDiff}
\end{figure}
One of the characteristics of the new gravity is that all spherical objects  contain induced black holes. 
In the context of polytrope models this presents the problem that the central value of the potential cannot be used, as in   the Lane-Emden equation. We can however impose the polytropic condition from  (\ref{eqn:PolyCond}) onto numerical solutions to iteratively solve the problem. Multiplying  (\ref{eqn:PolyCond}) by \(4 \pi r^2\) and integrating yields
\begin{equation}
M=\int_0^R 4 \pi r^2 \rho dr = \int_0^R 4 \pi r^2 \left(\frac{\Phi _R- \Phi)}{K (n+1)}\right)^n dr
\end{equation}
and then \begin{equation}
K = \frac{1}{(n+1)M^{1/n}} \left(  \int_0^R4\pi r^2 (\Phi_R - \Phi)^n dr \right)^{1/n}
\end{equation}
A new density distribution and K value can now be calculated from an initial density distribution by cycling through the following relations iteratively
\begin{eqnarray}
\Phi(r) =&\frac{-G}{(1-\frac{\alpha}{2})} \left( \frac{1}{r} \int_0^r 4 \pi s^2 \rho(s) ds+ \right. \nonumber \\
&\qquad \left. + \frac{1}{r^\frac{\alpha}{2}} \int_r^R 4 \pi s^{1+\frac{\alpha}{2}} \rho(s) ds \right)  \nonumber\\
K = & \frac{1}{(n+1)M^{1/n}} \left(  \int_0^R4\pi r^2 (\Phi_R - \Phi)^n dr \right)^{1/n} \nonumber \\
\rho(r) =& \left(\frac{\Phi_R - \Phi(r)}{K (n+1)}\right)^n
\label{eqn:equationset}\end{eqnarray}

\section{Polytropic Solar Models}
For the sun a polytrope model with \(n=3\) is known to give a good approximation to conditions in the solar core as compared with the Standard Solar Model \cite{KippBook}.  This is known as the Eddington Standard Model. The polytrope model does well in comparison with the Standard Solar Model \cite{BahcallSSM}. To test the calculation method, setting \(\alpha=0\) should reproduce the results from the Lane-Emden equation, which is based on Newtonian gravity.  The results of starting with a uniform density and then iteratively finding the solution agree with the values published by Chandrasekhar \cite{ChandBook}. The density distribution also matched numerical solutions produced in Mathematica to the Lane-Emden equation.

\begin{figure}
\hspace{30mm}\includegraphics[scale=0.6]{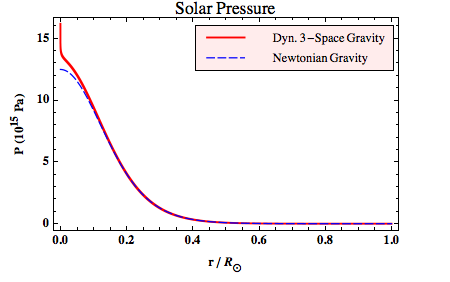} 

\hspace{30mm}\includegraphics[scale=0.6]{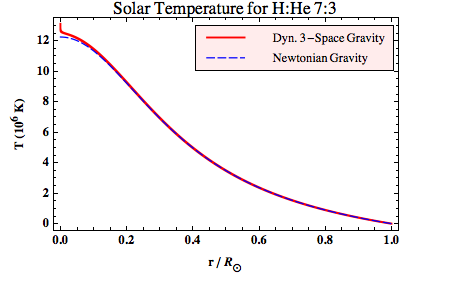} 
\caption{\small{The pressure and temperature in the center of the sun is predicted to be much larger in the new model.}}
\label{fig:PressureTemp}
\end{figure}

Results from solving  the equations in (\ref{eqn:equationset}) iteratively, until convergence was achieved, are shown in Figs.\ref{fig:GravityDensity}-\ref{fig:PressureTemp} for various quantities, and compared with the results for Newtonian gravity. 
For the new gravity (\(\alpha=1/137\)) we see a marked increase in the gravity strength $g(r)$ near the center, Fig.\ref{fig:GravityDensity}, caused by  the induced black hole at the center, which is characteristic of the new gravity theory, and which draws in the matter to enhance  the matter density near the center.  

The new model of gravity has been used to explain away the need for dark matter in astrophysics \cite{Paradigm}. Here we find the effective ``dark matter"  distribution that would need to be added to the matter distribution to create these gravitational effects in Newtonian Gravity. From (\ref{eqn:darkmatter}) and (\ref{eqn:PotentialDef}) we obtain
\begin{equation}
\rho_{DM}(r)=-\frac{\alpha}{8 \pi G r} \left(\frac{\Phi}{r} + \frac{d\Phi}{dr} \right)
\label{eqn:dmPhi1}\end{equation}
 Using  (\ref{eqn:RawGravitySol}) we then obtain
\begin{equation}
\rho_{DM}(r)=\frac{\alpha}{2} r^{-2-\alpha/2}\int_r^Rs^{1+\alpha/2}\rho(s)ds
\label{eqn:dmPhi2}\end{equation}
This effective ``dark matter" distribution is shown in Fig.\ref{fig:GravityDensity} for the polytropic sun model.
This then gives the total ``dark matter"
\begin{equation}
M_{DM}=\int_0^R4\pi r^2\rho_{DM}(r)dr=\frac{\alpha}{2} \frac{M}{1-\frac{\alpha}{2}}
\label{eqn:totalDM}\end{equation}
in agreement with (\ref{eqn:BHmass}). The ``dark matter" effect is  the same as the induced ``black hole" effect, in the new gravity theory.

The matter density has increased towards the center, as seen in Fig.\ref{fig:DensityDiff}, and so necessarily there is a slightly lower matter density in the inner middle region.  This effect is more clearly seen in the plot of $4\pi r^2 \rho(r)$.   The ``dark matter"/"black hole" effect contributes to the external gravitational acceleration, and so the total mass of the sun, defined as its  matter content,  is  lower than computed using Newtonian  gravity,  see (\ref{eqn:gamma}). The total mass is now  $0.37\%$  ($\equiv\alpha/(2-\alpha)$) smaller.

The pressure and temperature generated by the new gravity is shown in Fig.\ref{fig:PressureTemp}. The pressure comes from the polytrope relation, (\ref{eqn:Polytrope}),  and closely follows the density distribution. The temperature can be calculated from the ideal gas equation, with \(\mu=0.62\) corresponding to a ratio of 7:3 of Hydrogen to Helium, to obtain\begin{equation}
T(r)=\frac{P m_p \mu}{k\rho}
\end{equation}
where \(m_p\) is the mass of a proton, \(k\) is Boltzmann's constant and \(\mu\) is the mass ratio.
Unlike the pressure and density, the temperature is increased in the middle region as well as the inner region.

\section{Conclusions}

The discovery of the dynamical 3-space changes most of physics. This space has been repeatedly detected in light-speed anisotropy experiments.  The dynamics of this space follow from a unique generalisation of  Newtonian gravity, once that is expressed in a velocity framework.  Then the gravitational acceleration field ${\bf g}({\bf r},t)$ is explained as the local acceleration of the structured space, with evidence that the structure is fractal.  This space is the local absolute frame of reference.  Uniquely incorporating this space into a generalised Schr\"{o}dinger equation shows that, up to vorticity effects and relativistic effects, the quantum matter waves are refracted by the space, and yield that quantum matter has the same acceleration as that of space itself.
So this new physics provides a quantum theory derivation of the phenomenon of gravity.  The 3-space dynamics involves $G$ and the fine structure constant $\alpha$, with this identification emerging from the bore hole gravity anomalies, and from the masses of the minimal ``black holes" reported for globular clusters and spherical galaxies.  There are numerous other phenomena that are now accounted for, including a parameter-free account of the supernova red-shift - magnitude data.  The occurrence of $\alpha$ implies that we are seeing evidence of a new unified physics, where space and matter emerge from a deeper theory. One suggestion for this theory is {\it Process Physics}.

Herein we have reported the consequences of the new, emergent,  theory of gravity, when applied to the sun. This theory predicts  that the solar core,  which extends to approximately 0.24 of the radius,  is hotter, more dense and of higher pressure than current Newtonian-gravity based models.  Thus a new study is now needed on how these changes will   affect the solar neutrino output. It is also necessary to revisit the stellar evolution results.

\end{document}